\documentclass[a4paper,12pt]{article}
\usepackage[margin=1.05in]{geometry}
\usepackage[utf8]{inputenc}
\usepackage{graphicx}
\usepackage{subcaption}
\usepackage{color, colortbl}
\usepackage{amsmath}
\usepackage{physics}
\usepackage{setspace}
\usepackage{soul}
\usepackage{authblk}
\usepackage{color,soul}
\usepackage[square,numbers]{natbib}
\usepackage{pdfpages}

\newcommand{\specialcell}[2][c]{%
  \begin{tabular}[#1]{@{}c@{}}#2\end{tabular}}

\begin{document}
\title{Ferroelectricity and multiferroicity in anti-Ruddlesden-Popper structures}

\author[1]{Maxime Markov}
\author[1,3]{Louis Alaerts}
\author[1]{Henrique Pereira Coutada Miranda}
\author[1]{Guido Petretto}
\author[1]{Wei Chen}
\author[1]{Janine George}
\author[2]{Eric Bousquet}
\author[2]{Philippe Ghosez}
\author[1]{Gian-Marco Rignanese}
\author[1,3]{Geoffroy Hautier}
\affil[1]{UCLouvain,  Institute  of  Condensed  Matter  and  Nanosciences  (IMCN),Chemin des {\'Etoiles} 8,  B-1348  Louvain-la-Neuve,  Belgium}
\affil[2]{Theoretical Materials Physics, Q-MAT, CESAM, {Universit\'e} de {Li\`ege}, B-4000 Sart-Tilman, Belgium}
\affil[3]{Thayer School of Engineering, Dartmouth College, Hanover, New Hampshire 03755, USA}

\onehalfspacing
\maketitle

\begin{abstract}

Combining ferroelectricity with other properties such as visible light absorption or long-range magnetic order requires the discovery of new families of ferroelectric materials. Here, through the analysis of a high-throughput database of phonon band structures, we identify a new structural family of anti-Ruddlesden-Popper phases A$_4$X$_2$O (A=Ca, Sr, Ba, Eu, X=Sb, P, As, Bi) showing ferroelectric and anti-ferroelectric behaviors. The discovered ferroelectrics belong to the new class of hyperferroelectrics which polarize even under open-circuit boundary conditions. The polar distortion involves the movement of O anions against apical A cations and is driven by geometric effects resulting from internal chemical strains. Within this new structural family, we show that Eu$_4$Sb$_2$O combines coupled ferromagnetic and ferroelectric order at the same atomic site, a very rare occurrence in materials physics.

\end{abstract}

\section{Introduction}
Ferroelectric (FE) materials are of great fundamental and applied interests. They are currently used in many technologies such as electric capacitors, piezoelectric sensors and transducers, pyroelectric detectors, non-volatile memory devices, or energy converters~\cite{Lines:1977, Scott:1989, Scott:2007, Garcia:2009, Bowen:2014, Martin:2017,Kim:2018, Chanthbouala:2012}. For decades, most applications have relied on ferroelectric oxide perovskites. However, the need to combine ferroelectricity with other properties such as visible light absorption~\cite{Huang:2010,Li:2017} or long-range magnetic order~\cite{Spaldin:2019, Spaldin:2020} is driving the search for materials and structural classes beyond perovskites. High-throughput (HT) computational screening is a promising approach to search for materials meeting specific properties. It has been successfully used in a wide variety of fields from thermoelectrics~\cite{Chen:2016,Ricci:2017} to topological insulators~\cite{Li:2018,Zhang:2018,Choudhary:2019}. Different HT computing approaches have also been used to identify new ferroelectrics~\cite{Kroumova:2002,Bennett:2012,Garrity:2018, Smidt:2020}. Inspired by these previous studies and using a recently developed large phonon database, we searched for materials exhibiting dynamically unstable polar phonon modes, a signature of potential ferroelectricity. Our HT search identifies a new family of (anti-)ferroelectric materials: the series of anti-Ruddlesden-Popper phases of formula A$_4$X$_2$O with A a +2 alkali-earth or rare-earth element and X a $-$3 anion Bi, Sb, As and P. We survey how (anti-)ferroelectricity subtly depends on the chemistry of A$_4$X$_2$O and unveil the physical origin of the polar distortion. Interestingly, the discovered ferroelectrics belong to the new class of hyperferroelectrics~\cite{Garrity:2014} in which spontaneous polarization is maintained under open-circuit boundary conditions. The anti-Ruddlesden-Popper phases also lead to unique combinations of properties for instance a rare combination of ferroelectricity with ferromagnetism in Eu$_4$Sb$_2$O.

\section{Results}
\begin{figure}
\centering
    \includegraphics[width=0.8\textwidth]{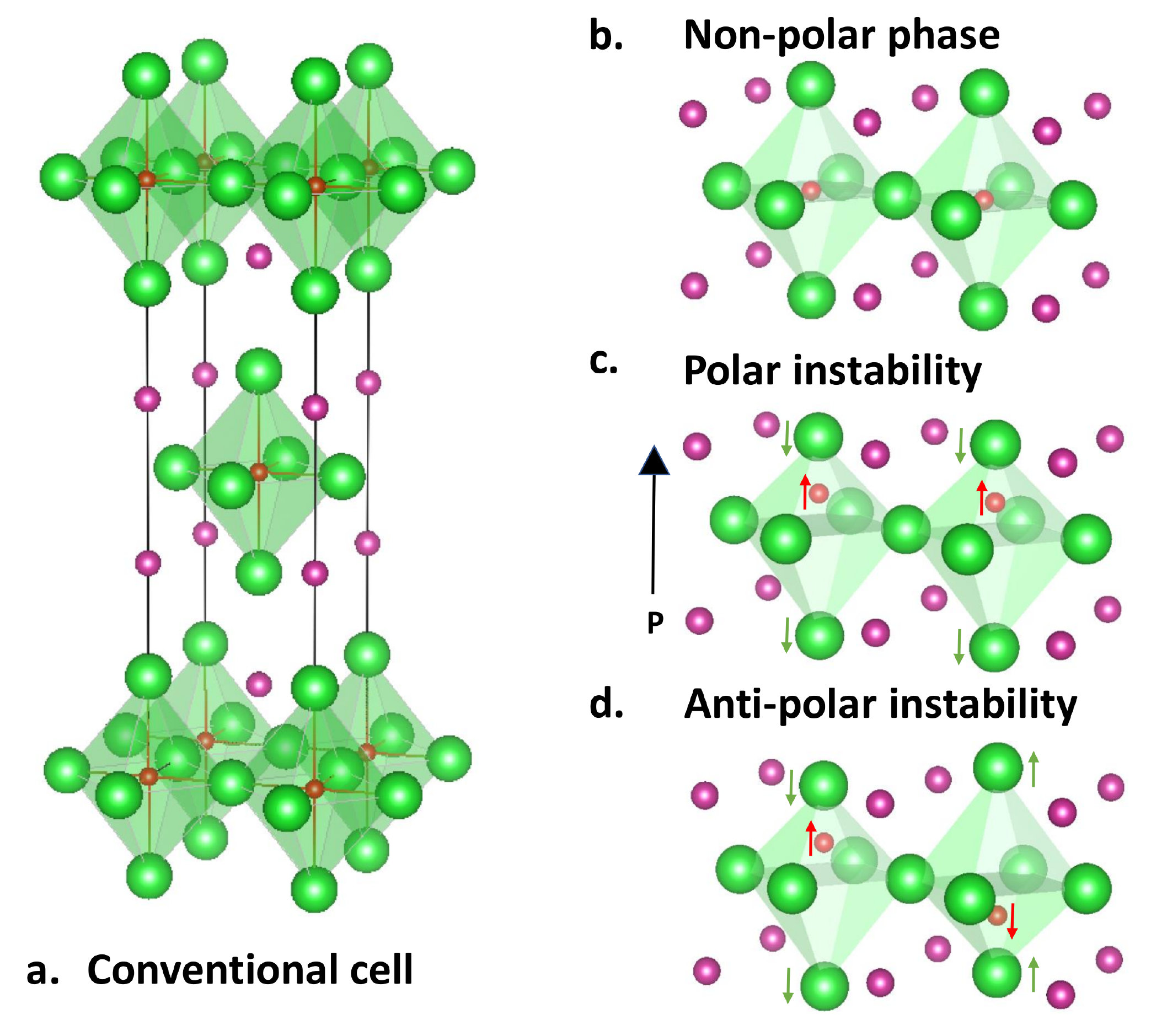}
   \caption{\label{fig:distrortions} (a) Conventional unit cell representing the anti-Ruddlesen-Popper structure of A$_4$X$_2$O. The A cation atoms (in green) form an octahedral cage with an O atom (in red) in its center. The X anion atoms (in violet) act as an environment in the voids surrounding the cages. Adopting a schematic representation with two neighboring octahedra surrounded by X atoms, we label three potentially metastable phases. In the 
   reference non-polar phase (b), the two O atoms are located in the middle of the octahedral cages of A cations (shaded green), being equidistant of the two apical A cations. Upon the polar distortion (c), the O atoms move upwards in the direction of apical A cations moving downwards as indicated by the red and green arrows respectively. This results in a loss of centrosymmetry and, thus, leads to a finite polarization value along this direction. In the case of 
   a anti-polar distortion (d), the O and A cation atoms in neighboring cages move in the opposite directions canceling out the polarization. In the plots, the displacements of the atoms have been amplified compared to their actual values (see text) in order to make them easily understood.}
\end{figure}

\begin{figure}
\center
  \begin{subfigure}[b]{0.5\textwidth}
    \includegraphics[width=\textwidth]{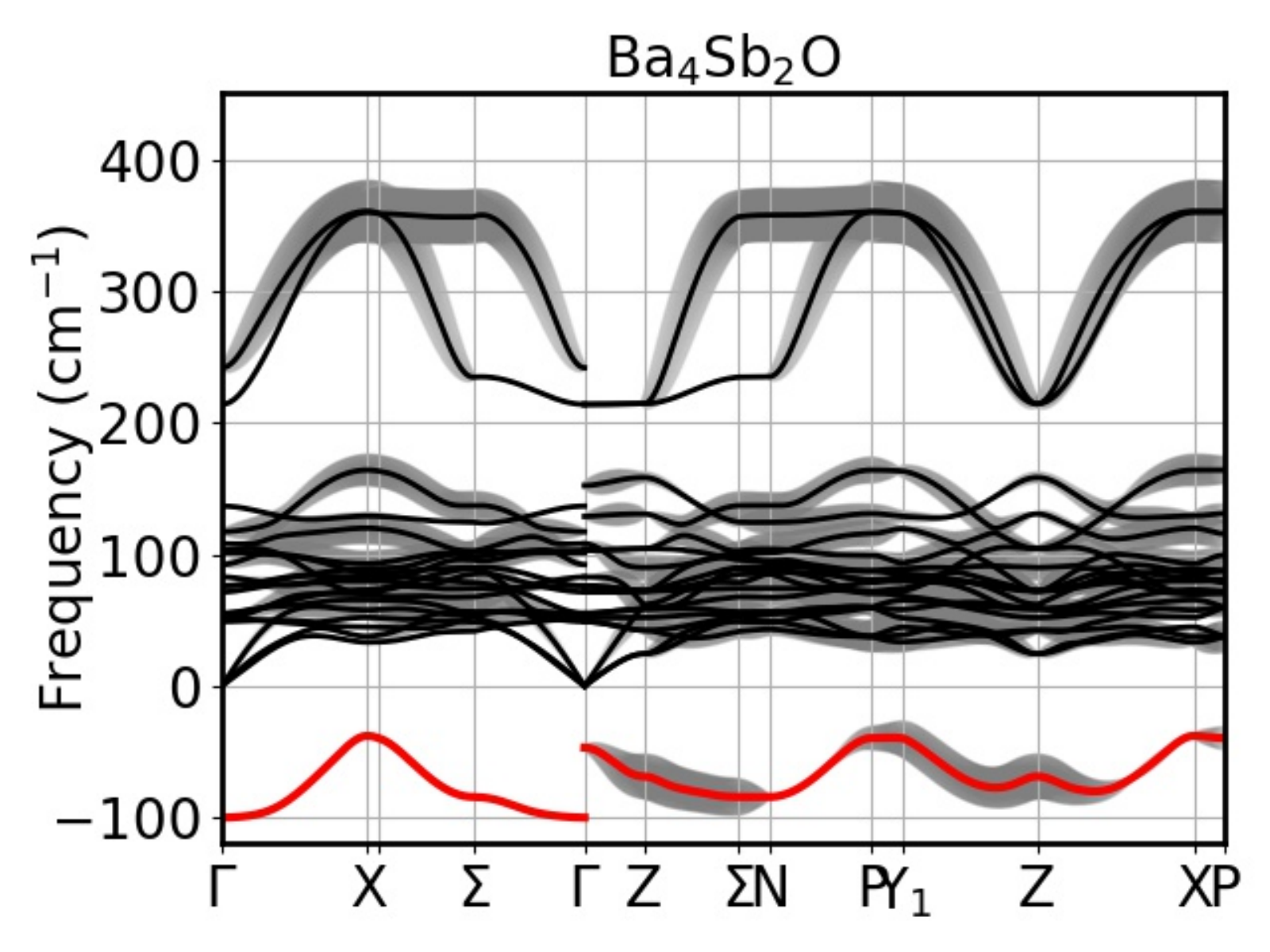}
    \label{fig:ph1}
  \end{subfigure}
  
  \begin{subfigure}[b]{0.5\textwidth}
    \includegraphics[width=\textwidth]{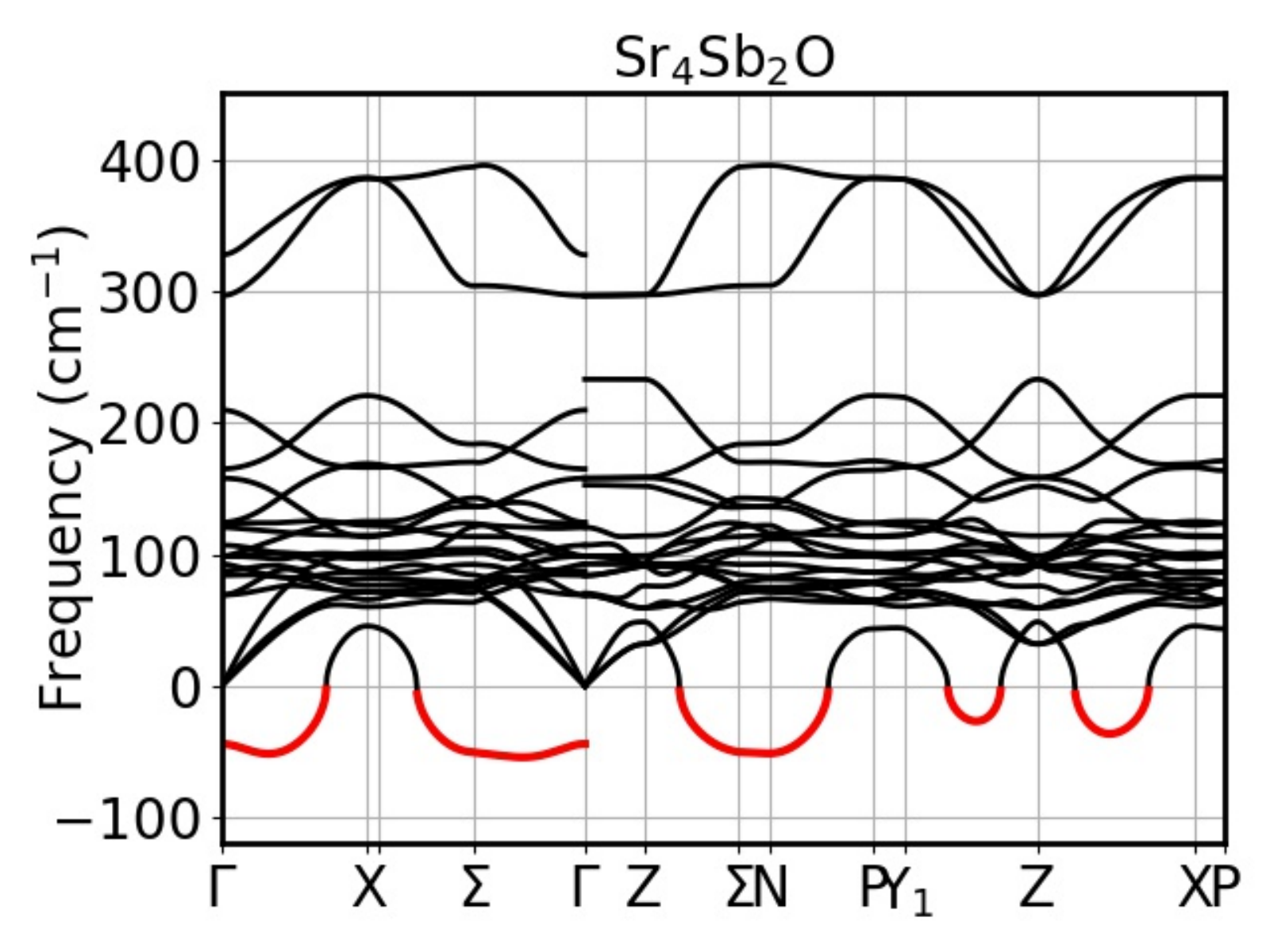}
    \label{fig:ph2}
  \end{subfigure}

  \begin{subfigure}[b]{0.5\textwidth}
    \includegraphics[width=\textwidth]{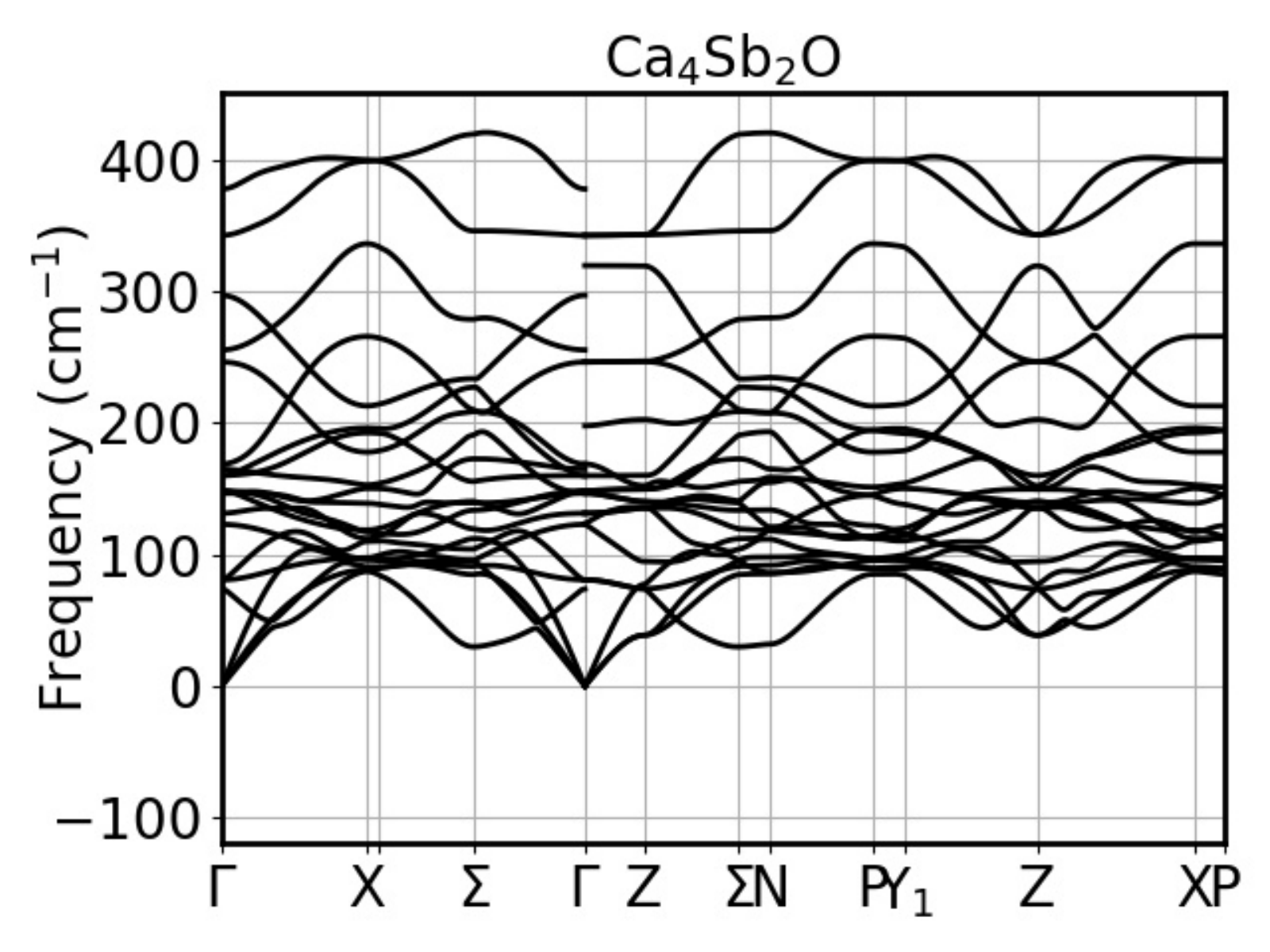}
    \label{fig:ph3}
  \end{subfigure}  
  \caption{\label{fig:phonons} Phonon dispersion curves of $I4/mmm$ A$_4$Sb$_2$O parent structures with the A cation atoms being Ba, Sr, Ca. Unstable phonon modes are highlighted in red. Change of the cation atom from the heavy Ba atom to the lighter Ca atom leads to the stabilization of the paralectric parent structure. On top of the phonon dispersion of Ba$_4$Sb$_2$O we plot the longitudinal character $L_(\mathbf{q},\nu)$ to distinguish between longitudinal and transverse optical modes and highlight a discontinuity at $\Gamma$.}
\end{figure}

A HT database of phonon band structures was recently built for more than 2,000 materials present in the Materials Project and mostly originating from the experimental Inorganic Crystal Structure (ICSD) database~\cite{Petretto:2018,Petretto:2018a,Jain:2013,Bergerhoff:1987,Zagorac:2019}. Using this database, we searched for non-polar structures presenting unstable phonon modes that could lead to a polar structure. This is the signature of a potential ferroelectric material~\cite{Garrity:2018}.
We identified Ba$_4$Sb$_2$O (space group $I4/mmm$) to be such a ferroelectric candidate. Its crystal structure and phonon band structure are shown in Figs.~\ref{fig:distrortions}a and~\ref{fig:phonons}a, respectively. This phase was reported experimentally by Röhr \textit{et al.}~\cite{Rohr:1996} and its crystal structure can be described as analogous to a Ruddlesden-Popper K$_2$NiF$_4$ phase, a naturally layered structure alternating rocksalt (KF) and perovskite (KNiF$_3$) layers, but for which cation and anions have been switched. Inspired by the terminology used for anti-perovskites~\cite{Krivovichev:2008,Bilal:2015}, we will refer to it as an anti-Ruddlesden-Popper phase. 

In Ba$_4$Sb$_2$O, the large instability of a polar phonon at $\Gamma$ is compatible with ferroelectricity. Relaxing the structure along this unstable mode confirms the existence of a lower-energy stable phase ($\Delta E = -6.58$~meV/atom) with a non-centrosymmetric space group $I4mm$ and a spontaneous polarization of 9.55 $\mu$C/cm$^{2}$. The parent $I4/mmm$ structure consists in the periodic repetition of alternative rocksalt BaSb and anti-perovskite Ba$_3$SbO layers, along what we will refer to as the $z$ direction. In this structure, O atoms are at the center of regular Ba octahedra (see Fig.~\ref{fig:distrortions}b). The polar distortion appearing in the $I4mm$ phase has an overlap of 90\% with the unstable polar mode. When keeping the center of mass of the system fixed, the related atomic displacement pattern, illustrated in Fig.~1c, is dominated by the movement along $z$ of O anion ($\eta_{O}=0.212$) against the apical Ba cations, that moves opposite way ($\eta_{O}=-0.029$)\footnote{In the $I4mm$ ground state, the motion of the top apical O atom has been reduced by anharmonic couplings with other modes.}. This cooperative movement of Ba and O atoms is responsible for the spontaneous polarization along $z$, while Sb and the other Ba atoms play a more negligible role (in reducing the polarization by only 4\%). Contrary to regular Ruddlesden-Popper compounds, that can show incipient in-plane ferroelectricity, the polarization is here along the stacking direction. Also, Ba$_4$Sb$_2$O does not show the antiferrodistortive instabilities ubiquitous in traditional Ruddlesden-Popper phases~\cite{Freedman:2009, Xu:2017b,Zhang:2017b}. 

Next to Ba$_4$Sb$_2$O, other alkali-earth atoms such as Ca and Sr have been reported to form in the same structure~\cite{Hadenfeldt:1991,Limartha:1980}. To further explore the role of chemistry on ferroelectricity, we plot in Fig.~\ref{fig:phonons} the phonon band structure of the  A$_4$Sb$_2$O series, with A = Ca, Sr and Ba, in their $I4/mmm$ phase. All compounds are insulating. We observe that the polar instability is reduced in Sr$_4$Sb$_2$O in comparison to Ba$_4$Sb$_2$O and is totally suppressed in Ca$_4$Sb$_2$O. 

The existence of a polar instability is not enough to guarantee a ferroelectric ground state. Other competing phases (e.g., anti-polar distortions) could be more stable than the polar phase. The presence of phonon instabilities at other points than $\Gamma$ (e.g., X or L) indicates the possibility for such competing phases (see Fig.~\ref{fig:phonons}). By following the eigendisplacements of individual and combined unstable modes, we confirm that the lowest energy phase is polar for Ba$_4$Sb$_2$O. Combined with its insulating character (HSE direct band gap is 1.22 eV) and the moderate energy difference between non-polar and polar states, this confirms a ferroelectric ground-state. In Sr$_4$Sb$_2$O, we find that the ground state is instead an anti-polar $C2/m$ phase as illustrated in Fig.~\ref{fig:distrortions}d (see Fig.~S2 for the entire crystal structure of the anti-polar distortion of Sr$_4$Sb$_2$O). This anti-polar phase is however only $\Delta E =$ 0.57~meV/atoms lower in energy than the polar phase. So, the polar phase could be stabilized under moderate electric fields, making Sr$_4$Sb$_2$O a potential antiferroelectric compound~\cite{Rabe:2013}. Using $E_c \approx \Delta E / \Omega_0 P_s$, we estimate the critical field $E_c$ in Sr$_4$Sb$_2$O  to be $81$ kV/cm, which could be easily accessible in experiment. Turning to the atomic pattern of anti-polar distortion, we see that it corresponds to a simple modulation of the polar distortion, with O atoms in neighboring octahedra moving in opposite directions and canceling out the macroscopic polarization (see Fig.~\ref{fig:distrortions}d). As such, Sr$_4$Sb$_2$O would therefore appear as a rare example of Kittel-type antiferroelectric~\cite{Kittel:1951,Rabe:2013,Milesi:2020}.

We note an intriguing discontinuity at $\Gamma$ in the unstable phonon branch of Ba$_4$Sb$_2$O (Fig.~\ref{fig:phonons}). We rationalize this discontinuity by noting that the unstable optical mode is polarized along the $z$ axis, so that it is transverse (TO) along $\Gamma$-X and $\Gamma$-Y and longitudinal (LO) along $\Gamma$-Z. This is further illustrated in Fig.~\ref{fig:phonons} by the grey smearing on top of the phonon dispersion curves that indicates the longitudinal character $L_{\mathbf{q},\nu}$. The latter was defined as $L_{\mathbf{q},\nu} = \frac{\mathbf{q}\cdot(Z^{*}_{\alpha}\cdot\mathbf{\Delta}_{\alpha,\nu})}{|Z^{*}_{\alpha}\cdot\mathbf{\Delta}_{\alpha,\nu}|}$, where $\mathbf{q}$ is the phonon wavevector, $Z^{*}$ is the Born effective charge matrix and $\mathbf{\Delta}$ is the eigen-displacement of atom $\alpha$ with phonon mode index $\nu$. Interestingly, we notice that the overlap between the lowest LO and TO mode eigendisplacements is of 90\% and that the LO-TO splitting is rather small so that the longitudinal mode remains strongly unstable.  Such a feature was previously reported in LiNbO$_3$~\cite{Veithen:2002}, or in hexagonal $ABC$ ferroelectrics and is the fingerprint of so-called hyperferroelectricity~\cite{Garrity:2014}. This demonstrates that Ba$_4$Sb$_2$O is not only ferroelectric but belongs to the interesting subclass of hyperferroelectrics in which a spontaneous polarization is maintained even under open-circuit boundary conditions (electrical boundary conditions with the electric displacement field $D=0$ ), so even when the unscreened depolarizing field tries to cancel out the bulk polarization.

\begin{table}[t]
\renewcommand{\arraystretch}{1.3}
\centering
\setlength\tabcolsep{3.5pt}
\begin{tabular}{|c|c|c|c|c|}
  \hline
       &               \textbf{Bi}             &  \textbf{Sb} & \textbf{As} & \textbf{P} \\
  \hline
    \textbf{Ba}   &  \specialcell{anti-ferroelectric \\ $C2/m$, -7.37~$\frac{\text{meV}}{\text{atom}}$} &  \specialcell{ferroelectric \\ $I4mm$, -6.58~$\frac{\text{meV}}{\text{atom}}$} &  \specialcell{ferroelectric \\ $I4mm$, -5.93$\frac{\text{meV}}{\text{atom}}$} & \specialcell{anti-ferroelectric \\ $Cmce$, -21.24$\frac{\text{meV}}{\text{atom}}$}\\
  \hline
    \textbf{Sr}   & \specialcell{anti-ferroelectric \\ $C2/m$, -1.74~$\frac{\text{meV}}{\text{atom}}$} & \specialcell{anti-ferroelectric \\ $C2/m$, -0.83~$\frac{\text{meV}}{\text{atom}}$} & \specialcell{paraelectric \\ $C2/m$, -0.65~$\frac{\text{meV}}{\text{atom}}$} & \specialcell{anti-ferroelectric \\ $Cmce$, -2.87~$\frac{\text{meV}}{\text{atom}}$} \\
  \hline
    \textbf{Ca}   & \specialcell{paraelectric \\ $I4/mmm$, 0.0~$\frac{\text{meV}}{\text{atom}}$} & \specialcell{paraelectric \\ $I4/mmm$, 0.0~$\frac{\text{meV}}{\text{atom}}$} & \specialcell{paraelectric \\ $I4/mmm$, 0.0~$\frac{\text{meV}}{\text{atom}}$} & \specialcell{paraelectric \\ $I4/mmm$, 0.0~$\frac{\text{meV}}{\text{atom}}$} \\
  \hline
\end{tabular}
\caption{\label{table:classification} Classification of the A$_4$X$_2$O family according to their electric state. Paraelectric refer to a stable structure or a structure with non-polar transition only, ferroelectric is a material with the non-polar to polar transition, anti-ferroelectric is a material with non-polar to non-polar transition with a polar phase being slightly higher in energy with respect to the lowest phase. The energy difference between the parent and the lowest child phase as well as the space group of the ground phase are shown. The parent phase has a space group $I4/mmm$, the polar and anti-polar phase space groups are $I4mm$ and $C2/m$ respectively. For A$_4$P$_2$O another orthorhombic anti-polar phase emerges.}
\end{table}

The chemical versatility of the anti-Ruddlesden-Popper phases is high. Beyond A$_4$Sb$_2$O oxo-antimonides, synthesis of oxo-phosphides, oxo-arsenates and oxo-bismuthides have been reported (see SI). We have systematically computationally explored the entire range of A$_4$X$_2$O structures (A=Ca, Sr, Ba; X=Sb, P, As, Bi). The phonon band structures are all plotted in Fig.~S1 and the results of the relaxation along all unstable phonon modes are presented in Table~\ref{table:classification}. More information on the phases competing for each chemistry is available in the SI. We found that all Ca-based compounds are paraelectric. Only Ba$_4$As$_2$O and Ba$_4$Sb$_2$O show a polar ground state. The ground states are most of the time anti-polar. We note that we only found few instabilities through octahedra rotations and tilts in the anti-Ruddlesden-Popper phase while they are common in standard Ruddlesden-Popper structures such as (Ca,Sr)$_3$Ti$_2$O$_7$~\cite{Zhang:2016a}, Ca$_3$Zr$_2$S$_7$~\cite{Zhang:2017}, La$_2$SrCr$2$O$_7$~\cite{Zhang:2016}. One of the appeal of perovskites is their strong chemical tunability as many different chemical substitutions can be performed tuning the ferroelectric properties~\cite{Benedek:2015, Zhang:2020}. It appears that similar tunability could be available for A$_4$X$_2$O. Moreover, as our described anti-Ruddlesden-Popper structure corresponds to $n=1$ in the traditional series A$_{3n+1}$X$_{n+1}$O$_{n+1}$, one could consider tuning properties by varying $n$ to higher values possibly by thin-film growth~\cite{Sharma:1998, Nie:2014}.

We now turn to the origin of the polar distortion in A$_4$X$_2$O. We especially focus on the A$_4$Sb$_2$O series which shows a transition in the nature of the ground state from strongly polar for Ba, to anti-polar for Sr and non-polar for Ca. The anti-Ruddlesen-Poppler structure shows a polar displacement of an anion in an octahedral cationic cage and it is natural to make the analogy with traditional ferroelectric perovskites such as BaTiO$_3$ where a cation moves in an anionic octahedral cage. However, the analysis of the Born effective charges hints at a very different physical mechanism in both situations. While ferroelectric oxide perovksites can show anomalously high Born effective charges ($Z^*_{Ti}$=+7.25 $e$, $Z^*_{0}$=$-$5.71 $e$)~\cite{Ghosez:1998}, the Born effective charges in Ba$_4$Sb$_2$O are closer to the nominal charges ($Z^*_{Ba}$=+2.67 $e$, $Z^*_{0}$=$-$2.71 $e$). This indicates a more ionic bonding between the O and alkali-earth atoms and that dynamical charge transfer is not as important as in oxide perovskites~\cite{Ghosez:1996}. This conclusion is further confirmed by the crystal orbital Hamilton population (COHP) analysis~\cite{Dronskowski:1993,Maintz:2016,Nelson:2020} showing rather weak ionic character of Ba-O bonds in Ba$_4$Sb$_2$O in contrast to the strong covalent character of Ti-O bonds in BaTiO$_3$ with ICOHP energy being one order of magnitude higher than the one in Ba$_4$Sb$_2$O. In passing, we note that while, Born effective charges are lower in anti-Ruddlesen-Popper structures, their large atomic displacements (e.g., 0.40 \AA\ for O and $-$0.20 \AA\ for one of Ba atoms in Ba$_4$Sb$_2$O) maintain a reasonable polarization. This analysis points out to a ferroelectric distortion driven by a geometrical effect with the simple picture of an O atom relatively free to move in a too large cationic cage. To further confirm this picture, we study the interatomic force constants (IFCs) in real space. We observe that the on-site IFC of the O atom, quantifying the restoring force that it feels when displaced with respect to the rest of the crystal, is close to zero in Ba$_4$X$_2$O along the $z$ (out-of-plane) direction, and one order of magnitude smaller than in-plane. This highlights that the O atoms are almost free to move along $z$ in the $I4/mmm$ phase. 

The close to nominal Born effective charges and very low on-site IFC are both characteristics of geometrically-driven ferroelectricity as described in fluoride perovskites~\cite{Garcia:2014}. The geometric nature of the instability naturally explains why going from Ba to Sr and Ca weakens the polar instability. Indeed, the on-site IFC of the O atom along $z$ increase as we go from Ba to Ca (0.17, 0.99, 1.79 eV/\AA$^2$) and as the cation to O distance along $z$ progressively decreases ($d_{AO} =$ 3.08, 2.88, 2.66 \AA). The smaller room for the O movement lowers the polar instability for Sr compared to Ba and cancels it for Ca. The local character of the structural instability in real space is confirmed by its fully delocalized character in reciprocal space in Fig.~2. The local and geometric nature of the structural instability is also consistent with the hyperferroelectric character~\cite{Khedidji:2020} and the possible emergence of antiferroelectricity.

\begin{figure}
\centering
\includegraphics[width=0.7\textwidth]{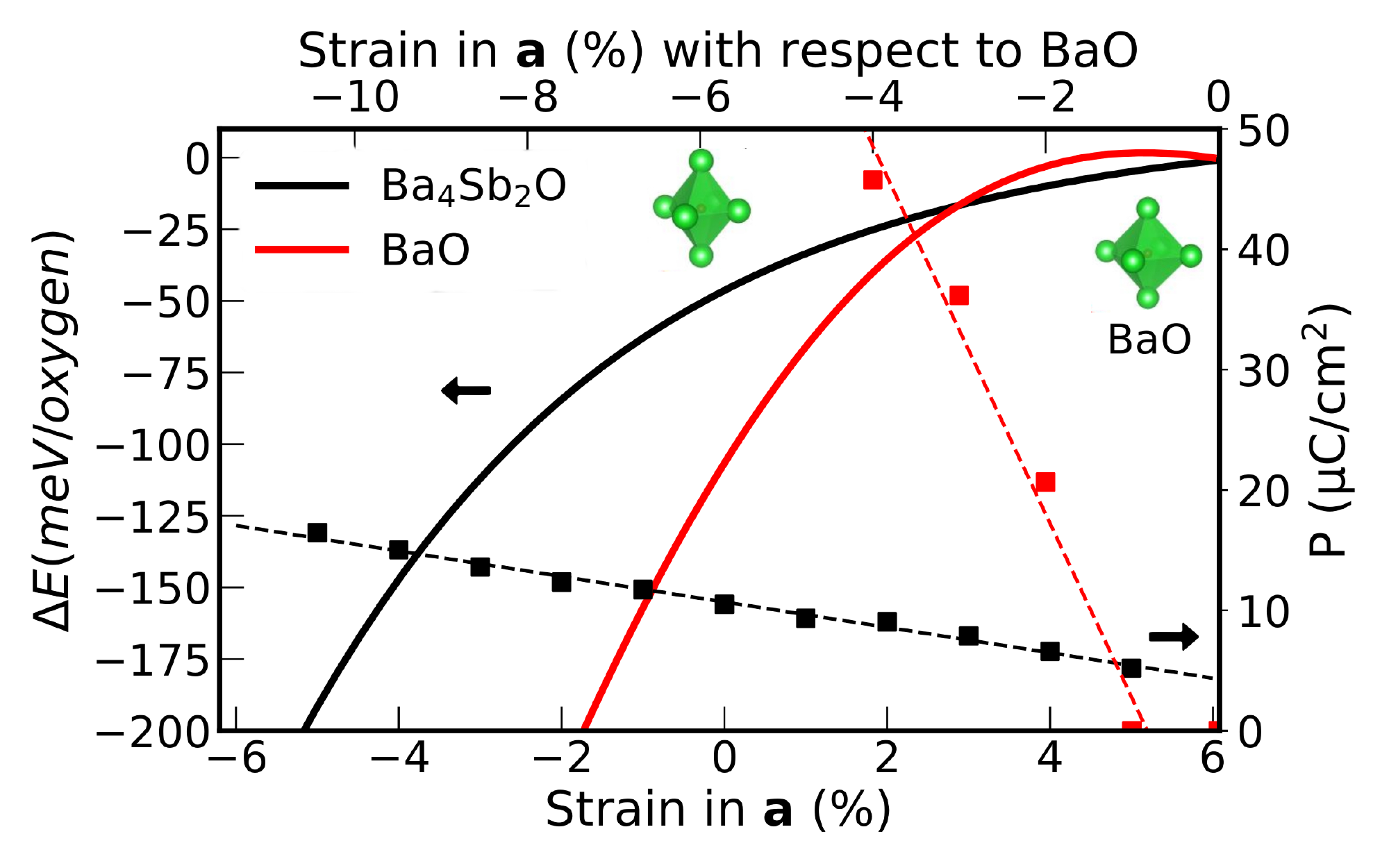}
  \caption{\label{fig:pdos} Energy difference between child and parent phases (solid lines) and polarization (dashed lines) of Ba$_4$Sb$_2$O (black curves) and BaO (red curves) as a function of in-plane strain computed with PBE functional. The data was fitted with linear function and the 4th order polynomials for polarization and energy difference respectively. Regular BaO and strained elongated Ba$_4$Sb$_2$O octahedra are shown.}
\end{figure}

In A$_4$X$_2$O anti-Ruddlesden-Popper compounds, the O atoms are surrounded by an octahedron of A atoms, showing a local environment similar to that experienced in the AO rocksalt phases. The latter are consistuted by regular octahedron units and are paraelectric. However, it has been predicted theoretically~\cite{Bousquet:2010} and recently confirmed experimentally~\cite{Goian:2020} that rock salt alkali-earth can become ferroelectric beyond a critical compressive epitaxial strain. Fig.~3 shows the energy difference and polarization between the paraelectric and ferroelectric phase as a function of the compressive strain for BaO (red) and Ba$_4$Sb$_2$O (black). The ferroelectric phase becomes favored for BaO above a compressive strain of 1\%. On the other hand, the unstrained Ba$_4$X$_2$O has Ba$_6$O octahedra distorted to the equivalent of around $-$6\% in BaO. Applying a tensile strain on Ba$_4$X$_2$O moves the octahedral geometry towards unstrained rock salt BaO and lowers the polar instability. Additionally, the $c/a$ ratio describing the octahedron elongation is $\sim$1.2 and close to that of ferroelectric phase of BaO at that strain. This highlights that in Ba$_4$X$_2$O, the surrounding atoms impose an internal, chemical strain on the Ba$_6$O cages. This natural strain induces ferroelectricity as previously highlighted in strained BaO. We note that such a level of strain (6\%) would be very difficult to reach within epitaxial films of rock salt. In Ba$_4$Sb$_2$O, polarization, however, increases much slower with strain than in BaO due to the presence of Sb atoms which limit the deformation of octahedra in the ferroelectric phase (see SI). While we focused on the X=Sb antimonide series here, the trend with Ba$>$Sr$>$Ca in terms of polar distortion is present across all chemistries from X=P, As, Bi and Sb (see SI).

\begin{figure}
\centering
    \includegraphics[width=0.6\textwidth]{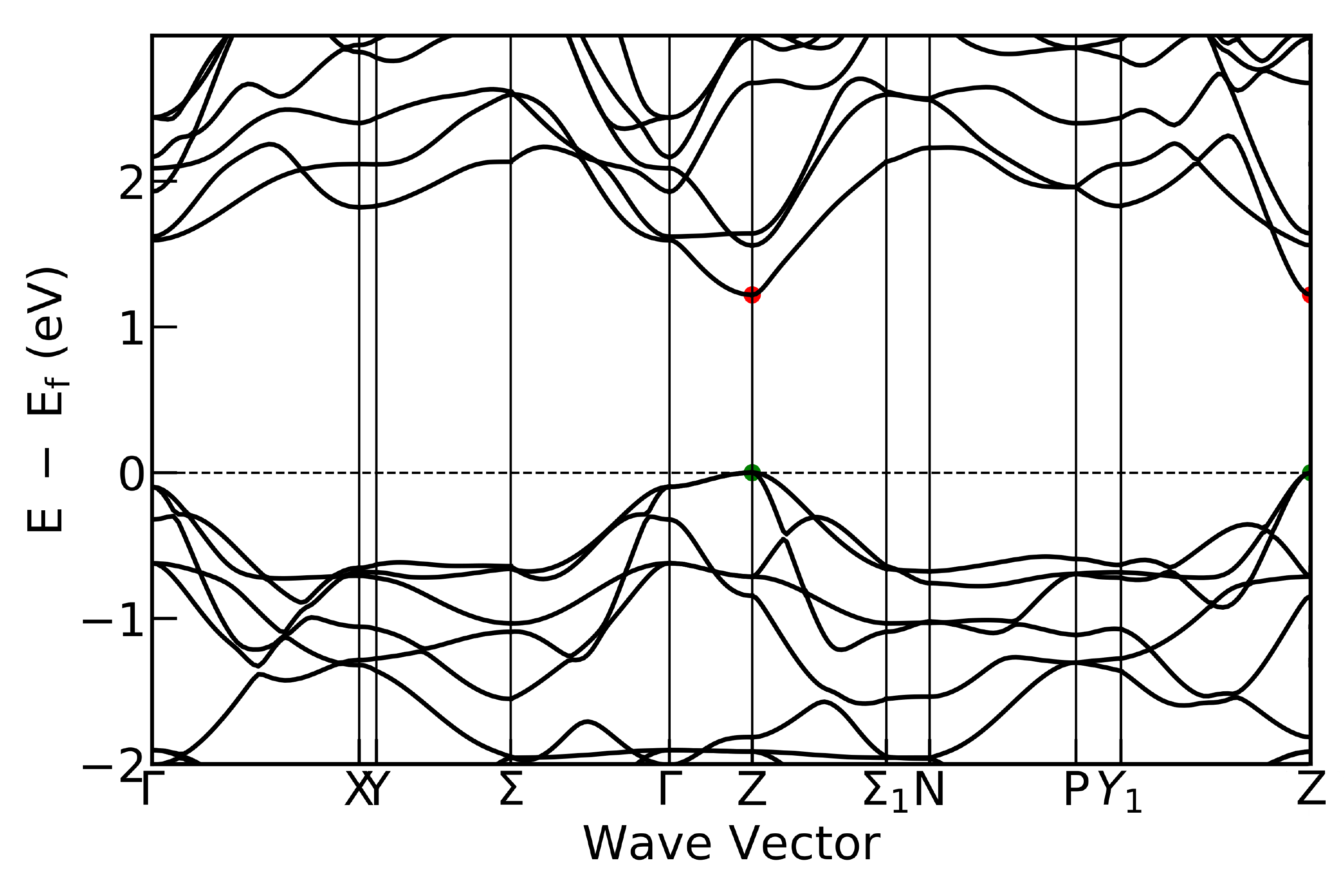}
  \caption{\label{fig:bandstructure} Electronic band structure of Ba$_4$Sb$_2$O in its $I4mm$ polar phase along the high-symmetry directions with PBE functional with a scissors correction of 0.55 eV. The direct band gap at Z point (1.22 eV) is marked by red and green points for the conduction and valence bands. }
\end{figure}

Compared to traditional perovskite-related structures, the A$_4$X$_2$O family offers opportunities in achieving properties that have been traditionally difficult to combine with traditional ferroelectric perovskites. Anti-Ruddlesen-Popper materials show typically smaller band gaps compared to oxide perovskites. While tetragonal $P4mm$ BaTiO$_3$ shows an indirect optical band gap of about 3.2 eV (1.67~eV in GGA between O 2p and Ti 3d states), we estimated the band gap of Ba$_4$Sb$_2$O to 1.22~eV using the HSE hybrid functional (0.67 eV in PBE). The band structure of Ba$_4$Sb$_2$O is shown in Fig.~\ref{fig:bandstructure}, highlighting a direct gap at Z between Ba 5d and O 2p states. Other A$_4$X$_2$O compounds show similar band gaps in the range from 0.57 to 1.00~eV in PBE (see Fig.~S4 in the SI). Such ferroelectrics with small band gaps compatible with visible light could be very useful in the field of ferroelectricity-driven photovoltaics~\cite{Huang:2010,Li:2017,Young:2012,Peng:2017,Grinberg:2013,He:2016}.

Another grand challenge has been to combine ferroelectricity with magnetic long-range order in magneto-electric multiferroics. The traditional mechanism of polar instability in the B site of a perovskite has been deemed difficult to combine with ferroelectricity since the non-magnetic d$^0$ character of the B site transition metal is often necessary to favor ferroelectricity~\cite{Hill:2000,Spaldin:2007}. Combining polar distortion on one site and magnetism on another site such as in EuTiO$_3$ or BiFeO$_3$~\cite{Shvartsman:2010,Spaldin:2019} or moving towards improper ferroelectricity as in YMnO$_3$ have lead to magnetoelectric multiferroics~\cite{Fennie:2005,Varignon:2013,Varignon:2019}. The geometrically driven polar instability demonstrated in anti-Ruddlesden-Popper structure offers an alternative opportunity for multiferroicity. Magnetic +2 rare-earth atoms often substitute to alkali-earth and Eu$_4$Sb$_2$O has been experimentally reported to form in the anti-Ruddlesden-Popper structure~\cite{Schaal:1998}. Computing phonon band structures and relaxing the structure along the unstable modes, we found that Eu$_4$Sb$_2$O is ferroelectric. Similar to Ba$_4$Sb$_2$O, the geometric polar instability in Eu$_4$Sb$_2$O involves directly the movement of non-magnetic O against the magnetic apical Eu$^{2+}$. This is likely to couple magnetism and ferroelectricity. Magnetic ordering computations show that Eu$_4$Sb$_2$O exhibits a ferromagnetic ground state with an easy axis pointing along the $c$ direction and along the polarization. We estimate the magnetic Curie temperature to be $\sim$24 K (see Methods). Most magnetoelectric multiferroic materials including the most studied BiFeO$_3$ are anti-ferromagnetic. Despite their technological importance, there are very few examples of materials combining ferromagnetic and ferroelectric order~\cite{Spaldin:2005} and the few known ones are double-perovskites (e.g., Pb$_2$CoWO$_6$~\cite{Brixel:1988} or the R$_2$NiMnO$_6$/La$_2$NiMnO$_6$ heterostructures~\cite{Zhao:2014}) where magnetism and ferroelectricity come from different sites. 
Eu$_4$Sb$_2$O as its parent rocksalt EuO is a ferromagnetic insulating oxide~\cite{Wei:2019}. 
The coexistence of ferromagnetism and ferroelectricity has just been confirmed experimentally in epitaxially strained EuO films~\cite{Goian:2020} and is naturally appearing in Eu$_4$Sb$_2$O anti-Ruddlesden-Popper phase. The magnetic space group $I4m'm'$ is compatible with linear magnetoelectric coupling and the magnetoelectric tensor has the following form~\cite{Gallego:2019}:
\begin{equation}
\alpha_{ME}  =
\begin{pmatrix}
 \alpha_{xx} & 0 & 0 \\
 0 &  \alpha_{xx} & 0 \\
 0 & 0 &  \alpha_{zz}
\end{pmatrix}
\end{equation}
More quantitatively, the computation of the linear magnetoelectric tensor in Eu$_4$Sb$_2$O confirms that a coupling is present with a non-negligible value: $\alpha_{xx} = 0.1$ ps/m (ionic contribution 0.08 ps/m and 0.02 electronic contribution), $\alpha_{zz}=0.016$ ps/m (ionic contribution 0.006 ps/m and 0.01 electronic contribution). We note that other rare-earth based anti-Ruddlesden-Popper phases are known to exist Eu$_4$As$_2$O~\cite{Wang:1977}, Eu$_4$Bi$_2$O~\cite{Honle:1998}, Yb$_4$As$_2$O~\cite{Burkhardt:1998}, Yb$_4$Sb$_2$O~\cite{Klos:2018} and Sm$_4$Bi$_2$O~\cite{Nuss:2011}. It is possible that in addition to Eu$_4$Sb$_2$O other anti-Ruddlesen-Popper compounds are magnetoelectric multiferroics.

\section{Conclusions}
Following a data-driven approach based on a HT search within a database of phonons, we have identified a family of A$_4$X$_2$O (A=Ba, Sr, Ca, Eu and X=Bi, Sb, As, P) materials forming in an anti-Ruddlesden-Popper structure and showing (anti-)ferroelectrics properties. The new mechanism of polar distortion involves the movement of an anion in a cation octahedron. This distortion is geometrically-driven and controlled by the natural strain present in the cation octahedron. This new mechanism leads to hyperferroelectricity but also offers the possibility to combine ferroelectricity with properties uncommon in traditional perovskite-based structures such as small band gaps or magnetism. More specifically, we show that Eu$_4$Sb$_2$O exhibits a rare combination of ferromagnetic and ferroelectric order coupled through linear magnetoelectric coupling. The wide range of chemistries forming in the anti-Ruddlesden-Popper structure offers a tunability similar to that of perovskite structures in terms of strain, chemistry and heterostructures and opens a new avenue for ferroelectrics research.

\section{Methods}

The high-throughput search for novel ferroelectrics was performed using a recently published phonon database~\cite{Petretto:2018}. We first selected the unstable materials presenting at least one phonon mode $m$ with imaginary frequencies $\omega_m(\mathbf{q})$ within a $\mathbf{q}$-point region of the Brillouin zone. For each of these materials and modes, we focused on the high-symmetry $\mathbf{q}$-points commensurate with a 2$\times$2$\times$2 supercell. We generated a set of new structures by moving the atoms in that supercell according to the displacements corresponding to the different modes and $\mathbf{q}$-points. The symmetry of each new structure was analyzed using the spglib library~\cite{Togo:2018} with a tolerance of $10^{-6}$~\AA\ and 1$^\circ$ on angles. Then, the new structures were categorized as polar or non-polar depending on their point group. Finally, after relaxing all the structures in the set, we classified the materials as paraelectric (when all the structures in the set are non-polar and the polarization is thus always zero), ferroelectric (when the ground state is polar hence possessing a finite polarization value), or anti-ferroelectric (when the ground state is non-polar but there exists at least one polar phase in the set slightly higher in energy). In the latter case, the material can be driven to the polar phase upon application of a strong enough electric field and thus acquire the non-zero polarization.

DFT calculations were performed with the ABINIT~\cite{Gonze:2020} and VASP~\cite{Kresse:1996,Kresse:1996a} codes. PBEsol exchange-correlation was used everywhere, if not otherwise noted. PseudoDojo norm-conserving scalar-relativistic pseudopotentials [ONCVSP v0.3]~\cite{Hamann:2013,vanSetten:2018} were used in ABINIT. The Brillouin zone was sampled using a density of approximately 1500 points per reciprocal atom. All the structures were relaxed with strict convergence criteria, i.e. until all the forces on the atoms were below $10^{-6}$~Ha/Bohr and the stresses below $10^{-4}$~Ha/Bohr$^3$~\cite{Petretto:2018}. The phonon bandstructures were computed within the DFPT formalism as implemented in ABINIT~\cite{Gonze:1997,Gonze:1997a} using a $\mathbf{q}$-point sampling density similar to the $\mathbf{k}$-point one though for $\Gamma$-centered grids. The polarization was computed with both the Berry-phase and Born effective charge approaches. GGA$\_$PBE PAW pseudopotentials were used in VASP~\cite{Kresse:1999}. The structures were relaxed up to $10^{-3}$~eV/\AA. The cut-off energy was set to 520~eV and electronic convergence was done up to $10^{-7}$~eV. The $\mathbf{k}$-point sampling was similar to the one used in ABINIT. Both codes yield essentially the same results in the identification of the ground state phase. The use of PBE exchange-correlation potential does not change the ground state phase as well. The Lobster calculations were performed based on VASP DFT calculations.~\cite{Dronskowski:1993,Maintz:2016,Nelson:2020} We used the following basis functions from the pbeVaspFit2015 for the projections:  Ca (3p, 3s, 4s), Sr (4p, 4s, 5s), Ba (5s, 5p, 6s),  Sb (5p, 5s), O (2p, 2s), Ti (3d, 3p, 4s). The \textbf{k}-point grids for these calculations were at least 12$\times$12$\times$3 for A$_4$X$_2$O and 13$\times$13$\times$13 for BaTiO$_3$.
 The magnetic structure calculations for Eu$_4$Sb$_2$O were performed with VASP code.
 The Eu pseudopotential includes 17 electrons in the valence. For the DFT+$U$ calculations, the parameters were set to $U$=6.0~eV and $J$=0.0~eV to accurately describe the localized Eu $f$ orbitals. Good electronic convergence up to $10^{-8}$~eV was obtained with an energy cut-off 600~eV and 6$\times$6$\times$3 $\mathbf{k}$-point grid. The results were double checked with a 12$\times$12$\times$6 $\mathbf{k}$-point grid. The Curie temperature was estimated using the random-phase approximation~\cite{Pajda:2001,Wei:2019}. The phonon bandstructure for Eu$_4$Sb$_2$O was computed through the finite displacements method as implemented in Phonopy~\cite{Togo:2018} using a 2$\times$2$\times$2 supercell. The electronic and ionic parts of magneto-electric tensor were computed with the magnetic field~\cite{Bousquet:2011} and finite displacements~\cite{Iniguez:2008} approaches, respectively. Magnetic symmetries and the form of magneto-electric tensor was identified via the Bilbao crystallographic server.

\section*{Acknowledgements}

This work was funded by the U.S. Department of Energy, Office of Science, Office of Basic Energy Sciences, Materials Sciences and Engineering Division under Contract No. DE-AC02-05-CH11231 : Materials Project program KC23MP. H.~P.~C.~M. acknowledges financial support from F.R.S.-FNRS through the PDR Grants HTBaSE (T.1071.15).
JG acknowledges funding from the European Union’s Horizon 2020 research and innovation program under the Marie Skłodowska-Curie grant agreement No 837910. The authors thank the Consortium des  Équipements de Calcul Intensif en Fédération Wallonie Bruxelles  (CÉCI) for computational  resources. Additionally,
the present research benefited from computational resources made available on the {Tier-1} 
supercomputer of the F\'ed\'eration Wallonie-Bruxelles, infrastructure funded by the Walloon 
Region under grant agreement n\textsuperscript{o}1117545.




\section*{Supplementary material}

# Supplementary material for "Ferroelectricity and multiferroicity through anion displacement in anti-Ruddlesen-Popper structures"

Maxime Markov[1], Louis Alaerts[1,3], Henrique Pereira Coutada Miranda[1], Guido Petretto[1], Wei Chen[1], Janine George[1], Eric Bousquet[2], Philippe Ghosez[2], Gian-Marco Rignanese[1], and Geoffroy Hautier[1,3]

[1] UCLouvain, Institute of Condensed Matter and Nanosciences (IMCN),Chemin des Étoiles 8, B-1348 Louvain-la-Neuve, Belgium
[2] Theoretical Materials Physics, Q-MAT, CESAM, Université de Liège, B-4000 Sart-Tilman, Belgium
[3] Thayer School of Engineering, Dartmouth College, Hanover, New Hampshire 03755, USA

# Contents

# S1 Material Synthesis

Anti-Ruddlesen-Popper $A_4X_2O$ phases have been synthesized in a wide range of chemistries. We list here the papers reporting the synthesis of these materials: $Ca_4P_2O$ [1, 2], $C_4As_2O$ [1], $C_4Sb_2O$ [3], $Sr_4P_2O$ [4], $Sr_4As_2O$ [4], $Ba_4P_2O$ [4], $Ba_4As_2O$ [4], $Yb_4As_2O$ [5], $Yb_4Sb_2O$ [6] and $Sm_4Bi_2O$ [7]. Another paper reports on the synthesis of a series of Anti-Ruddlesen-Popper phases to measure their possible superconductivity.[8] The magnetic compound $Eu_4Sb_2O$ studied in this paper has been reported in Ref. [9]. Moreover, other magnetic materials of the same type have also been synthesized: $Eu_4As_2O$ [10], $Eu_4Bi_2O$ [11]. The non-polar space group $I4/mmm$ (139) has been attributed to all of them. However, the latest results on the synthesis of $Ba_4Bi_2O$ [12] have shown that the O atoms prefer the coordination number 5 instead of 6, i.e. it moves from its central symmetric position in the cage towards the top/bottom cation atom in agreement with our theoretical results. Also, $Ba_4P_2O$ was found to have space group $Cmce$ (64) [4] which agrees well with our prediction for this material.

# S2 Competing phases

Fig.S1 shows the phonon band structures along the high-symmetry directions for all representative materials of $A_4X_2O$ family with $A = (Ba, Sr, Ca)$ atoms labeling the rows and $X = (Bi, Sb, As, P)$ labeling the columns. All band structures were computed with PBEsol exchange-correlation functional. The unstable imaginary modes are highlighted in red. The compounds with Ca are found to be stable and paraelectric. For the unstable materials with at least one phonon mode with imaginary frequency, we generated a set of new child structures by moving the atoms in a 2×2×2 supercell of parent structure according to the displacements corresponding to that mode and a high symmetry **q**-point (see the Methods section of the main text for details). The competing child phases and their energies relative to the energy of the phase are summarized in Table S1. The ground state of $Ba_4Sb_2O$ and $Ba_4As_2O$ are polar with space group $I4mm$. $Ba_4Bi_2O$, $Sr_4Bi_2O$, $Sr_4Sb_2O$ are anti-polar with the $C2/m$ ground state. $Ba_4P_2O$ and $Sr_4P_2O$ are anti-polar with space group $Cmce$. $Sr_4As_2O$ is paraelectric since it has only one non-polar child phase.

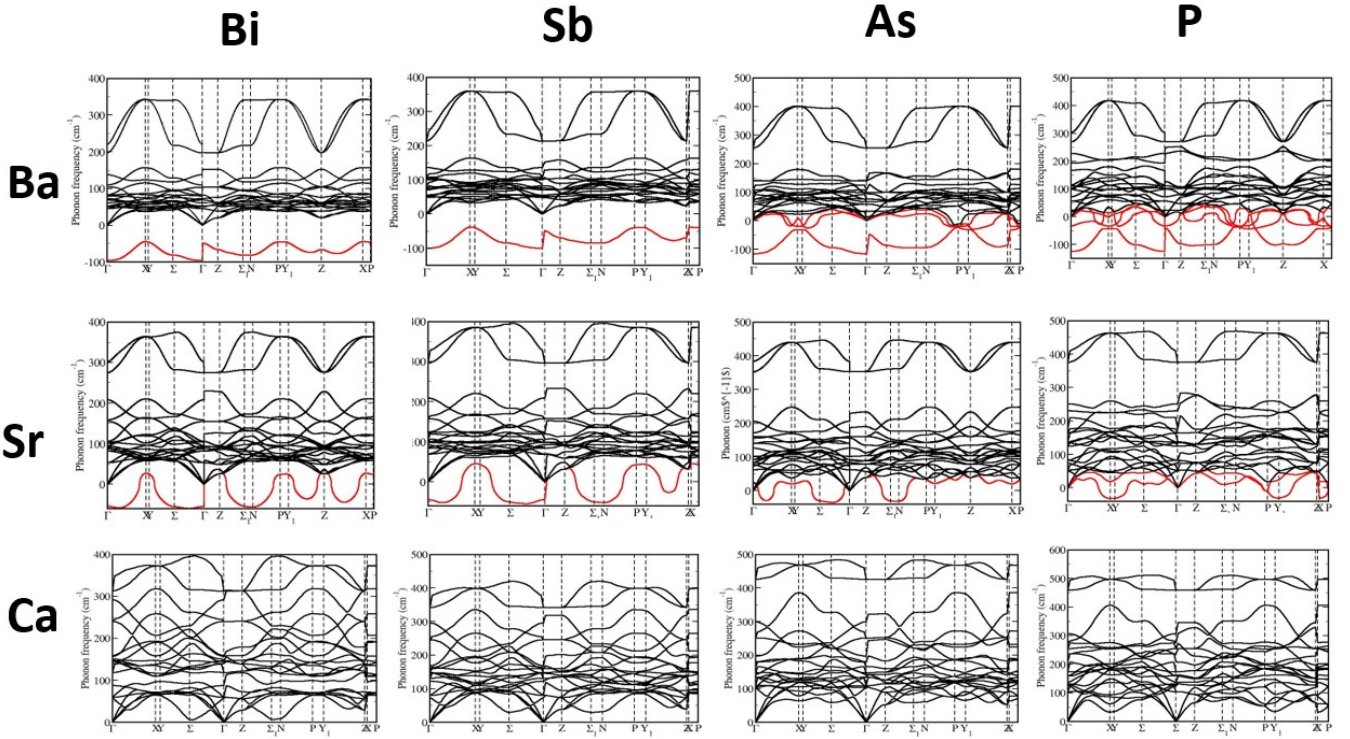

Figure S1. Phonon dipersions for the whole $A_4X_2O$ family of materials.

### Ba$_4$Bi$_2$O, no MPID

| crystal system | space group | class | q-point | $\Delta E^{tot}_{child-parent}$ meV/atom | $\omega$ cm$^{-1}$ |
|---|---|---|---|---|---|
| tetragonal | I4mm (107) | polar | $\Gamma$ | -7.215 | -96.08 |
| orthorhombic | Cmcm (63) | non-polar | X | -1.445 | -44.27 |
| tetragonal | P4/nmm (129) | non-polar | Z | -3.538 | -65.92 |
| **monoclinic** | **C2/m (12)** | **non-polar** | **N** | **-7.369** | -81.39 |

### Ba$_4$Sb$_2$O, mp-9774

| crystal system | space group | class | q-point | $\Delta E^{tot}_{ch-par}$ meV/atom | $\omega$ cm$^{-1}$ |
|---|---|---|---|---|---|
| **tetragonal** | **I4mm (107)** | **polar** | **$\Gamma$** | **-6.579** | -100.41 |
| orthorhombic | Cmcm (63) | non-polar | X | -0.837 | -38.10 |
| tetragonal | P4/nmm (129) | non-polar | Z | -2.841 | -68.98 |
| monoclinic | C2/m (12) | non-polar | N | -5.434 | -84.69 |

### Ba$_4$As$_2$O, mp-8300

| crystal system | space group | class | q-point | $\Delta E^{tot}_{child-parent}$ meV/atom | $\omega$ cm$^{-1}$ |
|---|---|---|---|---|---|
| **tetragonal** | **I4mm (107)** | **polar** | **$\Gamma$** | **-5.932** | -116.20 |
| orthorhombic | Cmce (64) | non-polar | X | -0.872 | -31.25 |
| tetragonal | P4/nmm (129) | non-polar | Z | -2.732 | -85.75 |
| monoclinic | C2/m (12) | non-polar | N | -4.862 | -93.64 |
| tetragonal | I4/mcm (140) | non-polar | P | -0.577 | -19.34 |

### Ba$_4$P$_2$O, no MPID

| crystal system | space group | class | q-point | $\Delta E^{tot}_{child-parent}$ meV/atom | $\omega$ cm$^{-1}$ |
|---|---|---|---|---|---|
| tetragonal | I4mm (107) | polar | $\Gamma$ | -6.07438 | -125.06 |
| orthorhombic | Imm2 (44) | polar | $\Gamma$ | -1.89806 | -33.13 |
| monoclinic | Cm(8) | polar | $\Gamma$ | -6.43405 | -33.13 |
| **orthorhombic** | **Cmce (64)** | **non-polar** | **X** | **-21.23915** | -44.24 |
| tetragonal | P4/nmm (129) | non-polar | Z | -2.96323 | -98.31 |
| monoclinic | C2/m (12) | non-polar | N | -4.75866 | -100.75 |

### Sr$_4$Bi$_2$O, mp-1025351

| crystal system | space group | class | q-point | $\Delta E^{tot}_{ch-par}$ meV/atom | $\omega$ cm$^{-1}$ |
|---|---|---|---|---|---|
| tetragonal | I4mm (107) | polar | $\Gamma$ | -1.10180 | -54.17 |
| **monoclinic** | **C2/m (12)** | **non-polar** | **N** | **-1.7378** | -55.56 |

### Sr$_4$Sb$_2$O, mp-755293

| crystal system | space group | class | q-point | $\Delta E^{tot}_{ch-par}$ meV/atom | $\omega$ cm$^{-1}$ |
|---|---|---|---|---|---|
| tetragonal | I4mm (107) | polar | $\Gamma$ | -0.2682 | -44.00 |
| **monoclinic** | **C2/m (12)** | **non-polar** | **N** | **-0.8379** | -51.24 |

### Sr$_4$As$_2$O, mp-8299

| crystal system | space group | class | **q**-point | $\Delta E^{tot}_{ch-par}$ meV/atom | $\omega$ cm$^{-1}$ |
|---|---|---|---|---|---|
| **monoclinic** | **C2/m (12)** | non-polar | N | **-0.64567** | -30.65 |

### Sr$_4$P$_2$O, mp-8298

| crystal system | space group | class | **q**-point | $\Delta E^{tot}_{ch-par}$ meV/atom | $\omega$ cm$^{-1}$ |
|---|---|---|---|---|---|
| monoclinic | C2 (5) | polar | P | -0.31355 | -9.69 |
| **orthorhombic** | **Cmce (64)** | non-polar | X | **-2.87213** | -31.14 |

### Eu$_4$Sb$_2$O, mp-1078192

| crystal system | space group | class | **q**-point | $\Delta E^{tot}_{ch-par}$ meV/atom | $\omega$ cm$^{-1}$ |
|---|---|---|---|---|---|
| **tetragonal** | **I4mm (107)** | polar | $\Gamma$ | **-2.01364** | -50.69 |
| monoclinic | C2/m (12) | non-polar | N | -0.07167 | -56.66 |

Table S1. Summary of the competing child phases for unstable parent structures $I4/mmm$

| material | A-O top | A-O side | top to side ratio |
|---|---|---|---|
| Ba$_4$Bi$_2$O | 3.10085 | 2.63494 | 1.176820 |
| Ba$_4$Sb$_2$O | 3.07889 | 2.61359 | 1.178031 |
| Ba$_4$As$_2$O | 2.99443 | 2.56257 | 1.168526 |
| Ba$_4$P$_2$O | 2.95853 | 2.54410 | 1.162898 |
| Sr$_4$Bi$_2$O | 2.89060 | 2.48617 | 1.162672 |
| Sr$_4$Sb$_2$O | 2.87443 | 2.46216 | 1.167442 |
| Sr$_4$As$_2$O | 2.80354 | 2.40124 | 1.167538 |
| Sr$_4$P$_2$O | 2.76835 | 2.37916 | 1.163583 |
| Ca$_4$Bi$_2$O | 2.65976 | 2.34607 | 1.133709 |
| Ca$_4$Sb$_2$O | 2.65670 | 2.31923 | 1.145510 |
| Ca$_4$As$_2$O | 2.61193 | 2.24849 | 1.161637 |
| Ca$_4$P$_2$O | 2.58113 | 2.22380 | 1.160684 |
| BaO | 2.76740 | 2.76740 | 1.000000 |
| SrO | 2.56744 | 2.56744 | 1.000000 |
| CaO | 2.38806 | 2.38806 | 1.000000 |

Table S2. Bond lengths (in Angstroms) between A-O (top) and A-O (side) s well as their ratio in parent phase of A$_4$X$_2$O-family and rocksalt binary AO oxides. Here A represents Ba,Sr,Ca and O stands for oxygen. A$_4$X$_2$O can be considered as rocksalt oxides chemically strained in the top direction.

## S3 Chemical strain. Analogy with rocksalt binary oxides.

The strain-induced ferroelectricity in rocksalt binary oxides such as BaO, SrO, CaO and EuO has been proposed in Ref. [13]. The strain is usually induced by epitaxy due to the lattice mismatch with a substrate. A$_4$X$_2$O can be considered as an alternative way to induce the strain which is now of chemical nature. In Table S2, we show the cation-O bond lengths in top and side directions of an octahedron. In rocksalt oxides, the ratio is always 1 i.e. an octahedron is symmetric. In A$_4$X$_2$O, the bonds along the top direction are elongated with respect to the side direction. Ba-compounds have larger bond ratio than other materials. This results in stronger ferroelectric behaviour similar to what

## S4 Non-polar $C2/m$ and $Cmce$ phases

Fig. S2 illustrates the projections of the conventional unit cell for anti-polar monoclinic $C2/m$ (left panel) and orthorhombic $Cmce$ (right panel) phases. The monoclinic distortion originates from the phonon eigenvector at the N point. It consists of the anti-phase movement of O atoms in the neighbor cages in contrast to the ferroelectric $I4mm$ distortion with the in-phase movement. The orthorhombic distortion originates from the eigenvector at the X point. It consists of the octahedra twisting in clockwise and anti-clockwise directions. In both phases, the polarization is canceled out because of the opposite distorions of the two sublattices. The twisting distortion becomes dominant only when the environment cation $X$ is phosphorous *i.e.* for $Ba_4P_2O$ and $Sr_4P_2O$. No distortions with twisting or rotations of octahedra has been observed for compounds with Bi, Sb.

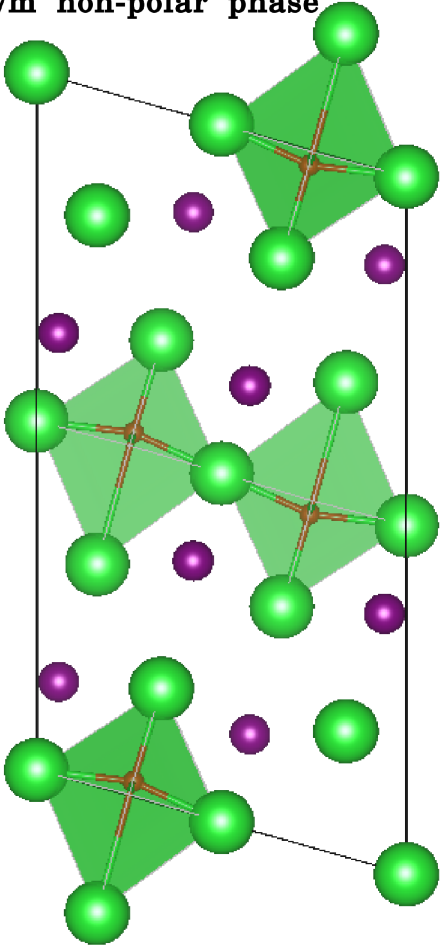
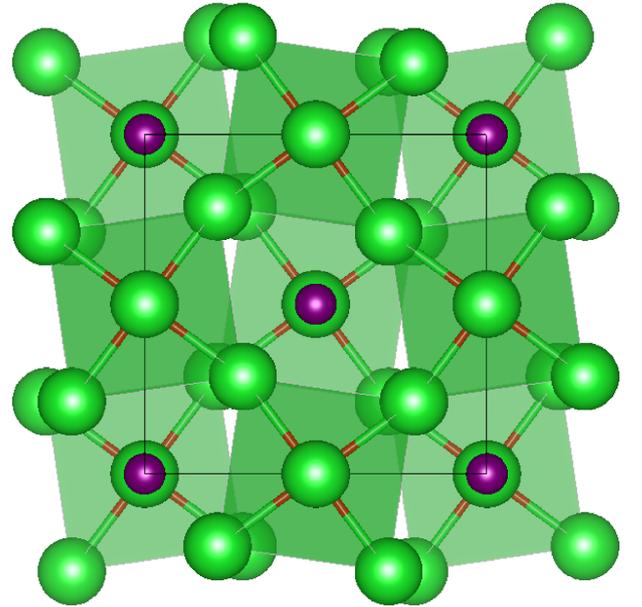

Figure S2. Projections of the conventional unit cells of anti-polar $C2/m$ and $Cmce$ phases. Left panel: monoclinic $C2/m$ phase. Cation atoms (green) form an octahedral cage with an O atom inside (red). The O atom is displaced from its center positiion and moves in the opposite direction in the neighbor cages making the structure centrosymmetric. The environment atoms of another $X$ cation are shown in violet. Right panel: orthorhombic $Cmce$ phase. O atoms rest in the center of the cage while the octahedra are twisted in opposite directions.

| Ba$_4$Sb$_2$O parent | | | | |
|---|---|---|---|---|
| atom | | $Z^*_{xx}$ | $Z^*_{yy}$ | $Z^*_{zz}$ |
| Ba$^{2+}$ | $\parallel$ | 3.15 | 3.15 | **2.67** |
| Ba$^{2+}$ | $\parallel$ | 3.15 | 3.15 | **2.67** |
| Ba$^{2+}$ | $\perp$ | 3.28 | 2.84 | **2.26** |
| Ba$^{2+}$ | $\perp$ | 2.84 | 3.28 | **2.26** |
| Sb$^{3-}$ | - | -4.44 | -4.44 | **-3.58** |
| Sb$^{3-}$ | - | -4.44 | -4.44 | **-3.58** |
| O$^{2-}$ | - | -3.55 | -3.55 | **-2.71** |

| Ba$_4$Sb$_2$O child | | | | |
|---|---|---|---|---|
| atom | | $Z^*_{xx}$ | $Z^*_{yy}$ | $Z^*_{zz}$ |
| Ba$^{2+}$ | $\parallel$, SB | 3.14 | 3.14 | **2.74** |
| Ba$^{2+}$ | $\parallel$, LB | 2.97 | 2.97 | **2.44** |
| Ba$^{2+}$ | $\perp$ | 3.25 | 2.76 | **2.43** |
| Ba$^{2+}$ | $\perp$ | 2.76 | 3.25 | **2.43** |
| Sb$^{3-}$ | - | -4.35 | -4.35 | **-3.45** |
| Sb$^{3-}$ | - | -4.27 | -4.27 | **-3.79** |
| O$^{2-}$ | - | -3.49 | -3.49 | **-2.80** |

| Sr$_4$Sb$_2$O parent | | | | |
|---|---|---|---|---|
| atom | | $Z^*_{xx}$ | $Z^*_{yy}$ | $Z^*_{zz}$ |
| Sr$^{2+}$ | $\parallel$ | 2.65 | 2.65 | **2.28** |
| Sr$^{2+}$ | $\parallel$ | 2.65 | 2.65 | **2.28** |
| Sr$^{2+}$ | $\perp$ | 2.89 | 2.16 | **2.17** |
| Sr$^{2+}$ | $\perp$ | 2.16 | 2.89 | **2.17** |
| Sb$^{3-}$ | - | -3.50 | -3.50 | **-3.09** |
| Sb$^{3-}$ | - | -3.50 | -3.50 | **-3.09** |
| O$^{2-}$ | - | -3.36 | -3.36 | **-2.73** |

| Sr$_4$Sb$_2$O child | | | | |
|---|---|---|---|---|
| atom | | $Z^*_{xx}$ | $Z^*_{yy}$ | $Z^*_{zz}$ |
| Sr$^{2+}$ | $\parallel$, SB | 2.66 | 2.66 | **2.32** |
| Sr$^{2+}$ | $\parallel$, LB | 2.62 | 2.62 | **2.21** |
| Sr$^{2+}$ | $\perp$ | 2.89 | 2.15 | **2.21** |
| Sr$^{2+}$ | $\perp$ | 2.15 | 2.89 | **2.21** |
| Sb$^{3-}$ | - | -3.49 | -3.49 | **-3.05** |
| Sb$^{3-}$ | - | -3.46 | -3.46 | **-3.17** |
| O$^{2-}$ | - | -3.35 | -3.35 | **-2.72** |

| Ca$_4$Sb$_2$O parent | | | | |
|---|---|---|---|---|
| atom | | $Z^*_{xx}$ | $Z^*_{yy}$ | $Z^*_{zz}$ |
| Ca$^{2+}$ | $\parallel$ | 2.66 | 2.66 | **2.19** |
| Ca$^{2+}$ | $\parallel$ | 2.66 | 2.66 | **2.19** |
| Ca$^{2+}$ | $\perp$ | 2.86 | 2.01 | **2.16** |
| Ca$^{2+}$ | $\perp$ | 2.01 | 2.86 | **2.16** |
| Sb$^{3-}$ | - | -3.38 | -3.38 | **-2.95** |
| Sb$^{3-}$ | - | -3.38 | -3.38 | **-2.95** |
| O$^{2-}$ | - | -3.41 | -3.41 | **-2.82** |

Table S3. Born effective charges of parent *I4/mmm* (left column) and child *I4mm* (right column) phases of A$_4$Sb$_2$O. We mark the cation atoms with $\perp$ and $\parallel$ symbols referring to atoms in-plane perpendicular to the polarization direction and parallel to it respectively. For the child structure, we distinguish between the atoms forming a short (SB) and long (LB) bond with O atoms. The polarization direction is highlighted in bold.

# S5 Born effective charges

In typical ferroelectric materials, like BaTiO$_3$, large Born effective charges highlight dynamical transfer of charge in line with reduced short-range forces and large dipole-dipole interactions compatible with polar instabilities [14]. For example, the excess charge of Ti atom reaches the value of -3.86 in BaTiO$_3$. The situation is different in fluoroperovskites [15] and here in A$_4$X$_2$O. In Table S3, we summarize the Born effective charges of parent (left column) and child (right column) structures of (Ba,Sr,Ca)$_4$Sb$_2$O. The charges $Z^*_{zz}$ along the polarization direction $\parallel$ are highlighted in bold. We see that the charges are close to their nominal values indicated on the atomic symbols. For example, the largest excess charges in Ba$_4$Sb$_2$O are -0.71 for O atom and +0.67 for Ba atoms along the polarization direction. We observe a strong reduction in the cation's Born effective charge from 2.67 in Ba$_\parallel$ to 2.28 in Sr$_\parallel$ and to 2.19 in Ca$_\parallel$ in the parent phase. Upon the transition, the Born effective charges for the long bond (LB) with O atoms are reduced down to 2.43 for Ba$_{\parallel,LB}$ and to 2.21 for Sr$_{\parallel,LB}$. For the short bond (SB), the Born effective charges increase slightly to 2.74 for Ba$_{\parallel,SB}$ and to 2.32 for Sr$_{\parallel,SB}$. Thus, the mechanism of ferroelectricity in this system is not the one of BaTiO$_3$.

# S6 Bonding analysis

We perform two types of bonding analysis using the valence bond model (see Table S4) and more advanced integrated crystal orbital Hamilton population (ICOHP) analysis (see Table S5). The latter method directly includes the hybridization effects and shows the covalent bond strength of the material. The computational details are given in the Methods section of the main text. We see that in $Ca_4Sb_2O$ bonds are much shorter and the absolute value of the ICOHP energy is higher with respect to other parent phases. This agrees well with the fact that $Ca_4Sb_2O$ is dynamically stable. After the phase transition in $Ba_4Sb_2O$, the bonds between cation and O atoms remain the same in plane, while out-of-plane one bond is enhanced and another is weakened after the phase transition. This is due to the fact that the O atom moves out-of plane from its symmetric position in the cage towards one of the cation atoms. The valence bond model confirms all above observations. Indeed, we see that the bond valences $s_{ij}$ of cation-O bonds in parent phase become weaker in-plane and stronger out-of-plane with the change of cation from Ba to Sr and Ca respectively. As a result, the parent structure stabilizes for $Ca_4X_2O$.

|    | Bi     | Sb     | As     | P      |
|----|--------|--------|--------|--------|
| Ba | -0.388 | -0.411 | -0.472 | -0.496 |
|    | -0.110 | -0.117 | -0.147 | -0.161 |
| Sr | -0.370 | -0.394 | -0.465 | -0.494 |
|    | -0.124 | -0.129 | -0.157 | -0.172 |
| Ca | -0.359 | -0.386 | -0.467 | -0.499 |
|    | -0.154 | -0.155 | -0.175 | -0.190 |

Table S4. Bond valences $s_{ij}$ of cation-O bonds (top values: cation atom is in-plane, bottom values: cation atom is out-of-plane) in parent structure of $A_4X_2O$ where A=(Ba,Sr,Ca) is the column index and X=(Bi, Sb, As, P) is the row index.

| | Bond lengths: | | | | ICOHP: | | |
|---|---|---|---|---|---|---|---|
| bond | parent | child, short | child, long | | parent | child, short | child, long |
| Ba-O | 3.08 | 2.70 | 3.81 | | -0.27 | -0.33 | -0.19 |
| Sr-O | 2.88 | 2.66 | 3.16 | | -0.36 | -0.36 | -0.32 |
| Ca-O | 2.66 | - | - | | -0.46 | - | - |
| Ti-O | 2.01 | 1.82 | 2.20 | | -3.26 | -5.42 | -1.90 |

Table S5. Bond lengths (in Å) and ICOHP energies (in eV) for cation-O bonds along the polarization direction for parent and child structures of $A_4Sb_2O$ and $BaTiO_3$. In the parent structure of $A_4Sb_2O$, the O atoms are located in positions equally distant from the two cation atoms. In the child structure of $A_4Sb_2O$, the O atoms are displaced in the polarization direction towards one of the cation forming a strong (short) bond and breaking another (long) bond. In the parent structure of $BaTiO_3$, Ti is located in the position equally distant from the O atoms. In the child structure of $BaTiO_3$, Ti is displaced in the polarisation direction towards one of the O atoms forming a much stronger (and much shorter) bond and weakening another (now longer) bond. One can clearly see that the Ti-O bonds are much more covalent than the Ba-O, Sr-O, Ca-O bonds and the change of the bond strengths from parent to child is also much more pronounced in $BaTiO_3$ than in $Ba_4Sb_2O$ indicating a very different physical mechanism for the polar distortion in both cases.

# S7 Interatomic force constants

To explain the origin of ferroelectric phase transition in our system we perform the on-site inter-atomic force constants (IFCs) analysis summarized in Table S6. The on-site IFCs characterize the force acting

on an atom when it is displaced from its equilibrium position with all other atoms being fixed. For example, large on-site IFCs describe a strong restoring force returning the displaced atom back to its equilibrium position. This situation corresponds to a stable structure. In turn, a small on-site IFCs tell us that an atom can move freely in a given direction without experiencing a strong force from other atoms. In other words, an atom moves because it is not closely packed and has a free space in a given direction. This describes so called geometric ferroelectricity. Indeed, the on-site IFCs of O atoms are one order of magnitude lower than all other IFCs in all ferroelectric compounds $Ba_4X_2O$. On the contrary, they are large in stable paraelectric compounds $Ca_4X_2O$. Decomposing the on-site IFCs of O atoms into short- and long-range terms (SR and LR respectively) in $Ba_4Sb_2O$, we find that they compensate each other IFC (SR) = -1.88 eV/ang$^2$ and IFC (LR) = +2.00 eV/ang$^2$. The large SR contribution is required for existence of hyperferroelectricity [16] which we indeed observed in $Ba_4Sb_2O$.

**$A_4Bi_2O$ parent**

| on-site IFC \ Cation | Ca | Sr | Ba |
|---|---|---|---|
| IFC cation in-plane | 4.7596 | 3.9190 | 2.6596 |
| IFC cation out-of-plane | 3.4011 | 3.1921 | 2.7781 |
| IFC Bi | 5.3047 | 4.6458 | 3.7936 |
| IFC O | **1.7938** | **0.9901** | **0.1739** |

**$A_4Sb_2O$ parent**

| on-site IFC \ Cation | Ca | Sr | Ba |
|---|---|---|---|
| IFC cation in-plane | 4.8625 | 3.9646 | 2.6314 |
| IFC cation out-of-plane | 3.8558 | 3.5983 | 3.1280 |
| IFC Sb | 5.8566 | 5.0870 | 4.1638 |
| IFC O | **1.8618** | **1.0261** | **0.1195** |

**$A_4As_2O$ parent**

| on-site IFC \ Cation | Ca | Sr | Ba |
|---|---|---|---|
| IFC cation in-plane | 4.4029 | 3.4690 | 2.0445 |
| IFC cation out-of-plane | 5.2551 | 4.7731 | 4.0346 |
| IFC As | 6.3716 | 5.3931 | 4.3611 |
| IFC O | **1.8978** | **1.0125** | **-0.0446** |

**$A_4P_2O$ parent**

| on-site IFC \ Cation | Ca | Sr | Ba |
|---|---|---|---|
| IFC cation in-plane | 4.2911 | 3.3175 | 1.8424 |
| IFC cation out-of-plane | 5.9994 | 5.3834 | 4.5049 |
| IFC P | 6.8157 | 5.7361 | 4.6419 |
| IFC O | **2.0202** | **1.0718** | **-0.0825** |

Table S6. On-site inter-atomic force constants (IFCs) in the $z$ direction in $A_4X_2O$ parent phases (eV/Å$^2$ units). The on-site IFCs of O atoms in Ba compounds are at least one order of magnitude lower than the ones of other atoms. This indicates a small energy cost to displace an individual O atom away from its high-symmetry position, and is consistent with the observed mechanism of phase transition originating mainly from the O atoms (see the main text). On the contrary, the on-site IFCs of O atoms in Ca compounds are much larger resulting in the stabilization of parent phases in these compounds.

# S8 Band structures of FE phase.

Electronic band structures of all polar $I4mm$ (107) phases are shown in Fig. S4. All materials show a small (PBE) band gaps of 0.57 eV for $Ba_4Bi_2O$, 0.65 eV for $Ba_4Sb_2O$, 0.67 eV for $Ba_4As_2O$, 0.68 eV for $Ba_4P_2O$, 0.89 eV for $Sr_4Bi_2O$ and 1.00 eV for $Sr_4Sb_2O$. HSE calculation corrects the bandgap for $Ba_4Sb_2O$ from 0.65 eV to 1.22 eV. All Ba compound have a direct band gap at Z point. $Sr_4Bi_2O$ also has a direct band gap but at $\Gamma$ point. $Sr_4Sb_2O$ is an indirect semiconductor with the valance band maximum at $\Gamma$ and conduction band minimum at $Z$.

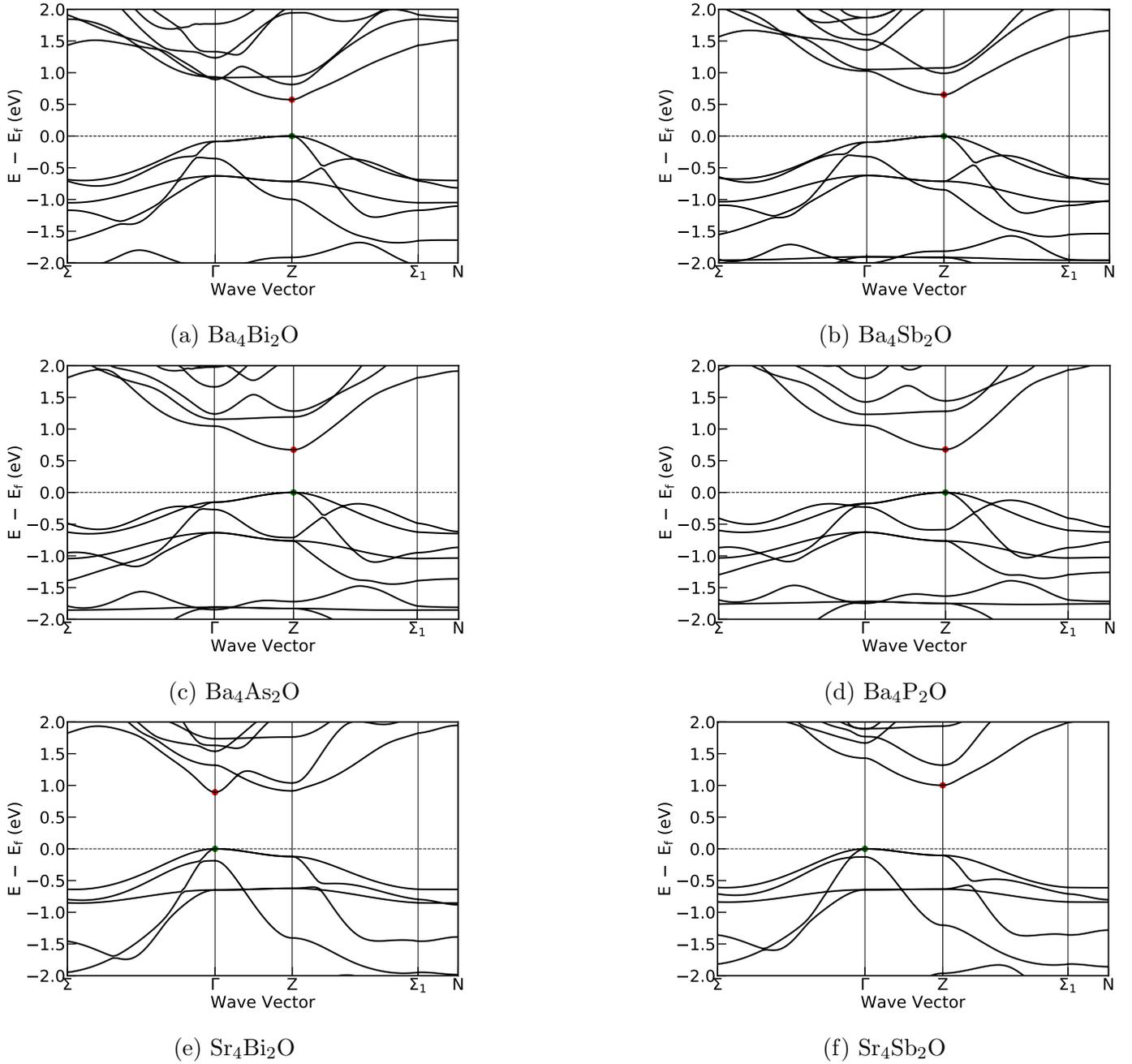

Figure S4. Electronic band structures of polar $I4mm$ phases along the high-symmetry directions computed with PBE exchange-correlation functional. Maximum top valence and minimum bottom conduction bands are highlighted with green and red points. All Ba compounds are direct gap semiconductors with the band gap at Z point. In $Sr_4Bi_2O$, the ban gap is at $\Gamma$ point. $Sr_4Bi_2O$ has an indirect band gap.

## S9 Effect of strain

In $A_4X_2O$ anti-Ruddlesden-Popper compounds, the O atoms are surrounded by an octahedron of A atoms, showing a local environment similar to that experienced in AO rocksalt phases. In the latter, it has been predicted theoretically [13] and recently confirmed experimentally [17] that a ferroelectric instability can emerge beyond a critical compressive epitaxial strain, that increases from BaO to SrO and CaO. In the $I4/mmm$ phase of $Ba_4X_2O$, the $Ba_6O$ octahedra is naturally strained compared to binary oxides, putting it well inside the ferroelectric regime. Moreover, the $c/a$ ratio describing the octahedron elongation is of about 1.2 and close to that of ferroelectric phase of BaO at that strain. This highlights that in $Ba_4X_2O$, the surrounding atoms impose an internal chemical strain on the

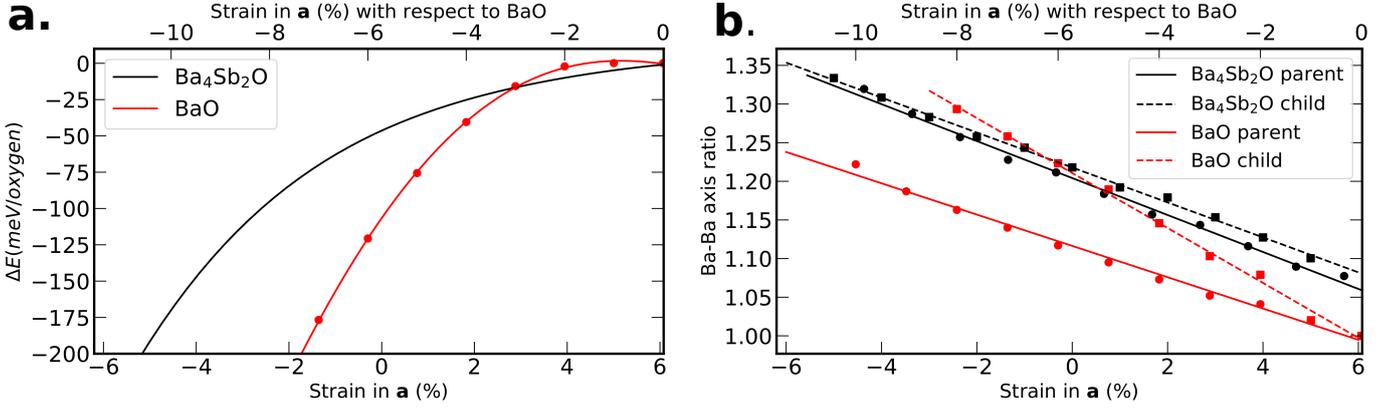

Figure S5. **Panel a**: energy difference between child and parent phases of $Ba_4Sb_2O$ (black curve) and BaO (red curve) as a function of in-plane strain computed with PBE functional. The data was fitted with the 4th order polynomial. **Panel b**: octahedron Ba-Ba axis ratio as a function of in-plane strain for parent (solid curves) and child (dashed curves) phases of $Ba_4Sb_2O$ (black curves) and BaO (red curves). The lines show a linear fit of raw data (circles and squares). The in-plane strain on both panels is given with respect to $Ba_4Sb_2O$ (bottom abscissa axis) and BaO (top abscissa axis) equilibrium lattice parameters.

$Ba_6O$ cages, that put the compound into a ferroelectric regime similar to that previously highlighted in BaO. This explanation is confirmed by similar trends in Fig. S5 that compares the evolution of the gain of energy between para- and ferro-electric phases (panel a) and $c/a$ ratio in $Ba_4Sb_2O$ and BaO (panel b) under epitaxial strain conditions. The evolution of $c/a$ further reveals that both $a$ and $c$ are constrained in $Ba_4X_2O$, which does not show significant polarization-strain coupling. This contributes to explain why $\Delta E$ evolves under strain less faster in $Ba_4X_2O$ than in BaO showing large coupling. Moving from Ba to Sr and Ca compounds, on the one hand the strain imposed to $A_3O$ cages in $A_4X_2O$ phases respect to $AO$ progressively decreases (from -5.6 to -4.1% and -2.9%) while, on the other hand, the critical strain required to produce ferroelectricity in related AO compounds increases (from -1.5% to -5.1% and -6.7%) [13].



%

\end{document}